\documentclass[pre,aps,twocolumn]{revtex4}
\usepackage{graphicx,epsf,subfigure,amsbsy,amssymb,amsmath,float}
\usepackage[latin1]{inputenc}

\newcommand{\ket}[1]{| #1 \rangle}

\newcommand{\bracket}[3]{\langle #1 | #2 | #3 \rangle}

\begin{document}

\title{Spin-orbit precession damping in transition metal ferromagnets}

\author{K. Gilmore$^{1,2}$, Y.U. Idzerda$^2$, and M.D. Stiles$^1$} 

\affiliation{$^1$ National Institute of Standards and Technology, Gaithersburg, 
MD 20899-6202 \\
$^2$ Physics Department, Montana State University, Bozeman, MT 59717}

\date{\today, {\it Journal of Applied Physics}}

\begin{abstract}

We provide a simple explanation, based on an effective field, for the precession 
damping rate due to the spin-orbit interaction.  Previous effective field treatments 
of spin-orbit damping include only variations of the state energies with respect to the 
magnetization direction, an effect referred to as the breathing Fermi surface.  
Treating the interaction of the rotating spins with the orbits as a perturbation, we 
include also changes in the state populations in the effective field.  In order to 
investigate the quantitative differences between the damping rates of iron, cobalt, and 
nickel, we compute the dependence of the damping rate on the density of states and the 
spin-orbit parameter.  There is a strong correlation between the density of states and 
the damping rate.  The intraband terms of the damping rate depend on the spin-orbit 
parameter cubed while the interband terms are proportional to the spin-orbit parameter 
squared.  However, the spectrum of band gaps is also an important quantity and does not 
appear to depend in a simple way on material parameters.

\end{abstract}


\maketitle


\section{Introduction}

Magnetic memory devices are useful if they can be reliably switched between two stable states.  The 
fidelity of this switching process depends sensitively on the damping rate of the system.  Despite 
decades of research and the relentless industrial push toward smaller and faster devices, many questions 
about the damping process remain unanswered, 
particularly for metallic ferromagnets.  Recent experimental efforts have investigated the extent to 
which the damping rate of NiFe alloys can be tuned through doping, particularly 
with the addition of rare earth \cite{Bailey:2001} and transition metal elements 
\cite{McMichael:2007}.  While these investigations found a general trend suggesting damping increases 
with increasing spin-orbit coupling of the dopant, the details behind this effect remained elusive.
To aid this effort, this article provides a simple description of the damping process 
and investigates how some material properties affect the damping rate.

Precession damping in metallic ferromagnets results predominantly from a combined 
effort of spin-orbit coupling and electron-lattice scattering \cite{Kambersky:1976, 
Gilmore:2007}.  The role of lattice scattering was studied in early experimental work through
the temperature dependence of damping rates \cite{Bhagat.Lubitz:1974, Heinrich:1966}.  
Measurement of 
damping rates versus temperature revealed two primary contributions to damping, an expected part that 
increased with temperature, and an unexpected part that decreased with temperature.  In 
cobalt these two opposing contributions combine to produce a 
minimum damping rate near 100 K, for nickel the increasing term is weaker leading 
to a temperature independent damping rate above 300 K, while for iron the 
damping rate becomes 
independent of temperature below room temperature.  Heinrich {\em et al}.~later noted that the 
temperature dependence of the increasing and decreasing contributions matched that of the 
resistivity and conductivity, respectively \cite{Heinrich:1979, Heinrich:1980}, and so dubbed the two 
contributions conductivity-like for the decreasing piece and resistivity-like for the increasing part.

Among the many theories on intrinsic precession damping \cite{Kambersky:1967, Kambersky:1970, 
Korenman.Prange:1972, Kambersky:1975, Kambersky:1976, Kambersky:2002, Choi:2004, 
Tserkovnyak:2004, MacDonald:2005, Heinrich.review:2005, Fahnle:2006, Bauer:2007}, Kambersk\'{y}'s 
torque-correlation model \cite{Kambersky:1976} is unique in qualitatively matching the observed 
non-monotonic temperature dependence that we just described.  We recently evaluated this model for iron, 
cobalt, and nickel, and showed that it accurately predicts the precession damping rates of 
these systems \cite{Gilmore:2007}.  While this model succeeds in capturing the important physical effects 
involved in precession damping, it does not easily identify the important physical processes or give 
insight into how one might alter the damping rate through sample manipulation.  In section II we briefly 
describe the torque-correlation model.  In order to provide a more tangible explanation of precession 
damping we rederive the damping rate from an effective field approach in section III.  This
discussion is followed in section IV by a quantitative analysis of the effect on the
damping rate of tuning the density of states and the spin-orbit parameter.

\section{Torque-correlation model}

Kambersk\'{y}'s theory describes damping in terms of 
the spin-orbit torque correlation function, finding a damping rate of
~
\begin{eqnarray}
\lambda &=& \pi \hbar \frac{\gamma^2}{\mu_0} \sum_{nm} \int \frac{d^3k}{(2\pi)^3} | 
\Gamma^-_{mn}(k) |^2 \nonumber \\
 &\times& \int d\epsilon_1 A_{nk}(\epsilon_1) A_{mk}(\epsilon_1) \eta(\epsilon_1) \, .
\label{lambda}
\end{eqnarray}

\noindent The gyromagnetic ratio is $\gamma = g\mu_0\mu_B/\hbar$, $g$ is the Land\'{e} $g$ 
factor, $\mu_0$ is the permeability 
of space, $n$ and $m$ are band indices, and ${\bf k}$ is the electron wavevector.  The matrix 
elements $|\Gamma^-_{mn}(k)|^2$ describe a torque between the spin and orbital moments that 
arises as the spins precess.  $\eta(\epsilon)$ is the derivative of the Fermi function                
$-df/d\epsilon$, which is a positive distribution peaked about the Fermi level that restricts 
scattering events to the neighborhood of the Fermi surface.  The electron spectral functions 
$A_{nk}(\epsilon)$ are Lorentzians in energy space centered at band energies with widths 
determined by the scattering rate.  They phenomenologically account for electron-lattice 
scattering.

Equation (\ref{lambda}) includes two processes: the decay of magnons into electron-hole 
pairs and the scattering of the electrons and holes with the lattice.  This expression is similar in 
structure to $sp$-$d$ models that have proven successful in describing dissipation in 
semiconductors \cite{MacDonald:2004}.  However, the physics of the magnon decay process is very 
different.  In the present case, there is no distinction between $sp$ and $d$ electrons.  The 
spin-orbit torque annihilates a uniform mode magnon and generates an 
electron-hole pair.  The electron-hole pair is then collapsed through lattice scattering.  The 
electron and hole are dressed through lattice interactions and are best thought of 
as a single quasiparticle with indeterminant energy and a lifetime given by the electron-lattice 
scattering time.  The dressed electron and hole can occupy the same band ($m=n$), which we call 
an {\em intraband} transition, or two different bands ($m \neq n$), an {\em interband} 
transition.  For intraband transitions, the integration over the spectral functions is 
proportional to the scattering time, just like the conductivity.  For interband transitions, the 
intregration over the spectral functions is roughly inversely proportional to the scattering 
time, as is the resistivity.  Therefore, the intraband terms in Eq.~(\ref{lambda}) give the 
conductivity-like contributions to damping that decrease with temperature while the interband 
terms yield the resistivity-like contributions that increase with temperature.

\section{Effective field derivation}

An effective field for the magnetization dynamics is defined as the variation of the electronic 
energy with respect to the magnetization direction $\mu_0 {\bf H}^{\rm eff} = - \partial E 
/\partial {\bf M}$.  The magnitude of the magnetization $M$ is considered constant within 
the Landau-Lifshitz formulation, only the direction $\hat{M}$ of the magnetization 
changes.  The total electronic energy of the system can be approximated by $E = \sum_{nk} \rho_{nk} 
\epsilon_{nk}$, which is a summation over the single electron energies $\epsilon_{nk}$ weighted by the 
state occupancies $\rho_{nk}$.  If the state occupancies are held at their equilbrium values, the 
resulting effective field is equivalent to that of the magnetocrystalline anisotropy \cite{Stiles:2001}, 
which describes reversible processes.  If however, the state occupancies are allowed to deviate from the 
equilibrium populations in response to the oscillating perturbation, an irreversible 
contribution also arrises, which we show produces the damping in Eq.~(\ref{lambda}).

As the magnetization precesses the energies of the states change through 
variations in the spin-orbit contribution and transitions between states occur.  These two 
effects, the changing energies of the states and the transitions between states, produce a 
contribution to the effective field
~
\begin{equation}
{\bf H}^{\rm eff} = -\frac{1}{\mu_0 M} \sum_{nk} \left [ \rho_{nk} \frac{\partial \epsilon_{nk}}{\partial
\hat{M}} + \frac{\partial \rho_{nk}}{\partial \hat{M}} \epsilon_{nk} \right ] \, .
\label{effH}
\end{equation}

\noindent The first term in the brackets describes the variation in the spin-orbit
energies of the states as the magnetization direction changes.  This effect, which has 
been discussed and evaluated before \cite{Kambersky:1970, Kambersky:2002, Fahnle:2005}, 
is generally referred to as the breathing Fermi surface model.  The spin-orbit torque does not cause 
transitions between states in this picture, but does cause the Fermi surface to
swell and contract as the magnetization precesses.  We will show that this portion
of the effective field gives the intraband terms of Eq.~(\ref{lambda}).  The second 
term in the brackets has previously been neglected in effective field treatments, but 
accounts for changes in the system energy due to transitions between states.  This term does not change 
the energies of the states, but does create electron-hole pairs by exciting electrons from lower bands to 
higher bands.  This process can be pictured as a bubbling of individual electrons on the Fermi surface.
We will demonstrate that this portion of the effective field gives the interband terms of 
Eq.~(\ref{lambda}).

\subsection{Intraband terms}

The first term in the effective field Eq.~(\ref{effH}) accounts for the effects of the 
breathing Fermi surface (bfs) model.  Since this model has previously been discussed in detail 
\cite{Kambersky:1970, Kambersky:2002, Fahnle:2005} we will give only a very brief review of 
it here, focusing instead on connecting it to the intraband terms of Eq.~(\ref{lambda}).

As the magnetization precesses the spin-orbit energy of each state changes.  Some occupied 
states originally just below the Fermi level get pushed above the Fermi level and simultaneously 
some unoccupied state originally above the Fermi level may be pushed below it.  This process takes 
the system, which was originally in the ground state, and drives it out of equilibrium into an 
excited state creating electron-hole pairs in the absence of any scattering events.  
Scattering, which occurs with a rate given by the inverse of the relaxation time 
$\tau$, brings the system to a new equilibrium.  The relaxation time approximation 
determines how far from equilibrium the system can get.
~
\begin{equation}
\rho_{nk} = f_{nk} - \tau \frac{df_{nk}}{dt} \, .
\label{rta}
\end{equation}

\noindent The occupancy $\rho_{nk}$ of each state $\psi_{nk}$ deviates from its equilibrium 
value $f_{nk}$ by an amount proportional to the scattering time.  How quickly the system 
damps depends on the magnitude of this deviation.  

The rate of change of the equilibrium distribution $df_{nk}/dt$ depends on how much the distribution
changes as the energy of the state changes $df_{nk}/d\epsilon_{nk}$, how much the state energy changes 
as the precession angle changes $d\epsilon_{nk}/d\hat{M}$, and how quickly the spin direction is 
precessing $d\hat{M}/dt$.  These can be combined with a chain rule
~
\begin{equation}
\frac{df_{nk}}{dt} = \frac{df_{nk}}{d\epsilon_{nk}} \frac{d\epsilon_{nk}}{d\hat{M}} 
\frac{d\hat{M}}{dt} \, .
\label{chain}
\end{equation}

\noindent Combining this result with the relaxation time approximation Eq.~(\ref{rta}) and substituting 
these state occupancies into the first term of the effective field in Eq.~(\ref{effH}) gives
~
\begin{eqnarray}
{\bf H}_{\rm bfs}^{\rm eff} &=& {\bf H}_{\rm bfs}^{\rm ani} + {\bf H}_{\rm bfs}^{\rm damp} \, ,\\
{\bf H}_{\rm bfs}^{\rm ani} &=& -\frac{1}{\mu_0 M} \sum_{nk} f_{nk} \frac{\partial 
\epsilon_{nk}}{\partial \hat{M}} \, , \\
{\bf H}_{\rm bfs}^{\rm damp} &=& -\frac{1}{\mu_0 M} \sum_{nk} \tau \left ( - 
\frac{df_{nk}}{d\epsilon_{nk}} \right ) \left ( \frac{d\epsilon_{nk}}{d\hat{M}} \right )^2 
\frac{d\hat{M}}{dt} \, .
\end{eqnarray}

\noindent ${\bf H}_{\rm bfs}^{\rm ani}$ is a contribution to the magnetocrystalline anisotropy field and 
${\bf H}_{\rm bfs}^{\rm damp}$ the damping field from the breathing Fermi surface model.  When we compare 
this damping field to the damping field postulated by the Landau-Lifshitz-Gilbert equation
~
\begin{equation}
{\bf H}_{\rm LLG}^{\rm damp} = -\frac{\lambda}{\gamma^2 M} \frac{d\hat{M}}{dt}
\label{LLG.effH}
\end{equation}

\noindent we find that the damping rate is
~
\begin{equation}
\lambda_{\rm bfs} = \tau \frac{\gamma^2}{\mu_0} \sum_{nk} \eta(\epsilon_{nk}) \left (
\frac{\partial \epsilon_{nk}}{\partial \hat{M}} \right )^2 \, .
\label{lambda_diag}
\end{equation}

\noindent As in Eq.~(\ref{lambda}), $\eta(\epsilon)$ is the negative derivative of the Fermi 
function and is a positive distribution peaked about the Fermi energy.

\begin{figure}
\includegraphics[angle=0,width=5.0cm]{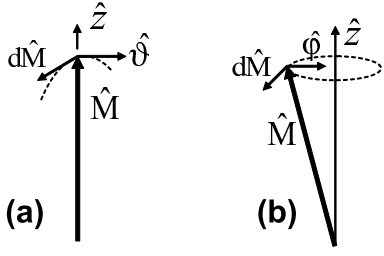}
\caption{Schematic description of precession geometry.  Within the breathing Fermi surface model (a) the 
damping rate is calculated as the magnetization passes through a specific point in a given direction.  
The torque correlation model (b) gives the damping rate for precessing about a given direction.  
Dashed curves indicate the precession trajectory.}
\label{schematic}
\end{figure}

As described in Fig.~(\ref{schematic}a), the result of the breathing Fermi surface model 
Eq.~(\ref{lambda_diag}) describes the damping rate of a 
material as the magnetization rotates through a particular point $\hat{z}$ about a given 
axis $\hat{\vartheta}$.  When $\hat{M}$ is instantaneously aligned with $\hat{z}$ the 
direction of the change in the magnetization $d\hat{M}$ will be perpendicular to $\hat{z}$, 
in the $\hat{x}$-$\hat{y}$ plane.  On the other hand, the torque correlation model 
Eq.~(\ref{lambda}) gives the damping rate when the magnetization is undergoing small angle 
precession about the $\hat{z}$ direction (see Fig.(\ref{schematic}b)).  When 
$\hat{z}$ is a high symmetry direction the change in the magnetization will stay in the 
$\hat{x}$-$\hat{y}$ plane.  In each scenario -- rotating $\hat{M}$ {\em through} $\hat{z}$ 
in the breathing Fermi surface model and rotating $\hat{M}$ {\em about} $\hat{z}$ in the 
torque correlation model -- $d\hat{M}$ is confined to the $\hat{x}$-$\hat{y}$ plane.  Therefore, 
rotating through $\hat{z}$ and rotating about $\hat{z}$ are equivalent in the small angle limit 
when $\hat{z}$ is a high symmetry direction.  With this observation we now show that the intraband 
contributions of the torque correlation model are equivalent to the breathing Fermi surface result 
under these conditions.

The only energy that changes as the magnetization rotates is the spin-orbit energy 
${\cal H}_{\rm so}$.  As the spin of the state $\ket{nk}$ rotates about the 
$\hat{\vartheta}$ direction by angle $\vartheta$ its spin-orbit energy is given by
~
\begin{equation}
\epsilon(\vartheta) = \bracket{nk}{e^{i\sigma \cdot \vec{\vartheta}} \, {\cal H}_{\rm so} \, 
e^{-i\sigma \cdot \vec{\vartheta}}}{nk} 
\end{equation}

\noindent where $\vec{\vartheta} = \vartheta \, \hat{\vartheta}$.  Taking the derivative of this energy 
with respect to $\vartheta$ in the limit that $\vartheta$ goes to zero shows that the energy derivatives 
are
~
\begin{equation}
\frac{\partial \epsilon}{\partial \vartheta} =  i\bracket{nk}{[\sigma \cdot \hat{\vartheta} \, , \, {\cal 
H}_{\rm so}]}{nk} \, .
\end{equation}

\noindent Figure (\ref{schematic}) shows that the derivative $\partial 
\epsilon/\partial \vartheta$ is identical to $\partial \epsilon/\partial \hat{M}$ and 
that when $\hat{M} = \hat{z}$ the rotation direction 
$\hat{\vartheta}$ lies in the $x-y$ plane.  The two components of the transverse torque operator 
$\Gamma^x$ and $\Gamma^y$ can be obtained (up to factors of $i$) by setting $\hat{\vartheta}$ equal to 
$\hat{x}$ or $\hat{y}$, respectively.  From this observation we find
~
\begin{equation}
|\bracket{nk}{\Gamma^-}{nk}|^2 = \left ( \frac{\partial \epsilon}{\partial x} \right )^2 + 
\left ( \frac{\partial \epsilon}{\partial y} \right )^2 \, .
\end{equation}

\noindent  When the magnetization direction $\hat{z}$ is pointed along a high symmetry 
direction the transverse directions $\hat{x}$ and $\hat{y}$ are equivalent and $|\Gamma^-|^2 = 
2(\partial \epsilon/\partial \hat{M})^2$.

Substituting the torque matrix elements for the energy derivatives in 
Eq.(\ref{lambda_diag}) gives a damping rate of 
~               
\begin{equation}
\lambda_{\rm bfs} = \frac{\tau \gamma^2}{2\mu_0} \sum_{nk} \left | \Gamma_{n}^-(k) 
\right |^2 \eta(\epsilon_{nk}) \, .
\end{equation}

\noindent For the intraband terms in Eq.~(\ref{lambda}) the integration over 
the spectral functions reduces to $\tau \eta(\epsilon_{nk})/2\pi\hbar$ so we find
~
\begin{eqnarray}
\lambda_{\rm bfs} &=& \pi \hbar \frac{\gamma^2}{\mu_0} \sum_{n} \int \frac{d^3k}{(2\pi)^3} 
\left | \Gamma_{n}^-(k) \right |^2 \nonumber \\
 &\times& \int d\epsilon_1 A_{nk}(\epsilon_1) A_{nk}(\epsilon_1) \eta(\epsilon_{1}) \, ,
\end{eqnarray}

\noindent which matches the intraband terms of Eq.~(\ref{lambda}).

\subsection{Interband terms}

As the magnetization precesses, the spins rotate and the spin-orbit energy changes.  This 
variation acts as a time dependent perturbation
~                           
\begin{equation}
V(t) = e^{i\sigma \cdot \varphi(t)} {\cal H}_{so} e^{-i\sigma \cdot \varphi(t)} - 
{\cal H}_{so}(0) \approx i[\sigma \cdot \varphi(t) , {\cal H}_{so}] \, .
\label{perturbation}
\end{equation}

\noindent This approximation results from linearizing the exponents, which is appropriate in the 
small angle limit.  The time dependence of the spin direction is $\hat{\varphi}(t) = \cos \omega t 
\, \hat{x} + \sin \omega t \, \hat{y}$, up to a phase factor.  This 
perturbation causes band transitions between the states 
$\psi_{nk}$ and $\psi_{mk}$.  The initial and final states have the same wavevector because 
these transitions are caused by the uniform precession, which has a wavevector of zero.  
The transition rate between states due to this perturbation is
~
\begin{equation}
W_{mn}(k) = \frac{2\pi}{\hbar} \left | \Gamma^-_{mn}(k) \right |^2 \delta(\epsilon_{mk} -
\epsilon_{nk} - \hbar\omega) \, .
\label{golden.rule}
\end{equation}

The variations of the occupancies of the states with respect to the 
magnetization direction are given by the master equation
~
\begin{equation}
\frac{\partial\rho_{nk}}{\partial t} = \sum_{m \neq n} W_{mn}(k) [\rho_{mk} - 
\rho_{nk}] \, .
\label{master}
\end{equation}

\noindent The second term in the effective field Eq.~(\ref{effH}) contains the factor $\partial 
\rho_{nk}/\partial\hat{M}$ which is $(\partial \rho_{nk}/\partial t) / (\partial \varphi/\partial t)$ 
where $\partial \varphi/\partial t = \omega$.  Inserting these expressions into the second term in the 
effective field and rearranging the sums gives
~
\begin{equation}
{\bf H}^{\rm eff} = -\frac{1}{2\mu_0 M} \sum_{nk} \sum_{m \neq n} \frac{W_{mn}(k)}{\omega^2} 
[\rho_{nk} - \rho_{mk}] [\epsilon_{mk} - \epsilon_{nk}] \frac{d\hat{M}}{dt} \, .
\end{equation}

\noindent Comparing this result to the effective field predicted by the Landau-Lifshitz-Gilbert 
equation (\ref{LLG.effH}) we find a damping rate of
~
\begin{equation}
\lambda = \frac{\gamma^2}{2\mu_0} \sum_{nk} \sum_{m \neq n} W_{mn}(k) \frac{[\rho_{nk} - 
\rho_{mk}]}{\omega} \frac{[\epsilon_{mk} - \epsilon_{nk}]}{\omega} \, .
\end{equation}

\noindent The finite lifetime of the states is introduced with the spectral functions
~
\begin{eqnarray}
\lambda &=& \frac{\hbar^2 \gamma^2}{2\mu_0} \sum_{nk} \sum_{m \neq n} \int d\epsilon_1 \,
A_{nk}(\epsilon_1) \int d\epsilon_2 \, A_{mk}(\epsilon_2) \, \nonumber \\
 &\times& W_{mn}(k) \frac{[f(\epsilon_1) - f(\epsilon_2)]}{\hbar\omega} \frac{[\epsilon_{2} 
- \epsilon_{1}]}{\hbar\omega} \, .
\end{eqnarray}

\noindent Inserting the transition rate Eq.~(\ref{golden.rule}), integrating over 
$\epsilon_2$, and taking the limit that $\omega$ goes to zero leaves
~
\begin{eqnarray}
\lambda &=& \pi\hbar\frac{\gamma^2}{\mu_0} \sum_{n} \sum_{m \neq n} \int \frac{d^3k}{(2\pi)^3} 
\left | \Gamma_{mn}^-(k) \right |^2 \nonumber \\
 &\times& \int d\epsilon_1 A_{nk}(\epsilon_1) A_{mk}(\epsilon_1) \eta(\epsilon_1) \, ,
\end{eqnarray}

\noindent which are the interband terms of Eq.~(\ref{lambda}).

In this derivation of the bubbling Fermi surface contribution to the damping we have ignored an 
additional, reversible term that contributes to the magnetocrystalline anisotropy.  
This contribution arises from changes in the equilibrium state occupancies as the magnetization 
direction changes.  This contribution to the magnetocrystalline anisotropy is localized to the Fermi 
surface while the contribution dicussed in IIIA is spread over all of the occupied levels.

\section{Tuning the damping rate}

We have previously demonstrated that the mechansim of thetorque correlation model 
Eq.~(\ref{lambda}) accounts for the majority of the precession damping rates of the transition 
metals iron, cobalt, and 
nickel \cite{Gilmore:2007}.  In the present work we have so far shown that this expression for 
the damping rate can be described simply within an effective field picture.  We now investigate 
the degree to which the damping rate may be modified by adjusting certain material 
parameters.  Inspection of Eq.~(\ref{lambda}) reveals that the damping rate depends on the 
convolution of two factors: the torque matrix elements and the integral over the spectral 
functions.  We separate the quantitative analysis of the damping rates into their 
dependencies on these two factors, beginning with the spectral weight.

The calculations for the damping rate of Eq.~(\ref{lambda}) discussed below are performed using the 
linear augmented planewave method in the local spin density approximation.  The details of the 
computational technique may be found in \cite{Gilmore:2007}, \cite{Stiles:2001}, and the included 
references.

\subsection{Spectral overlap}

\begin{figure}
   \includegraphics[angle=0,width=8.0cm]{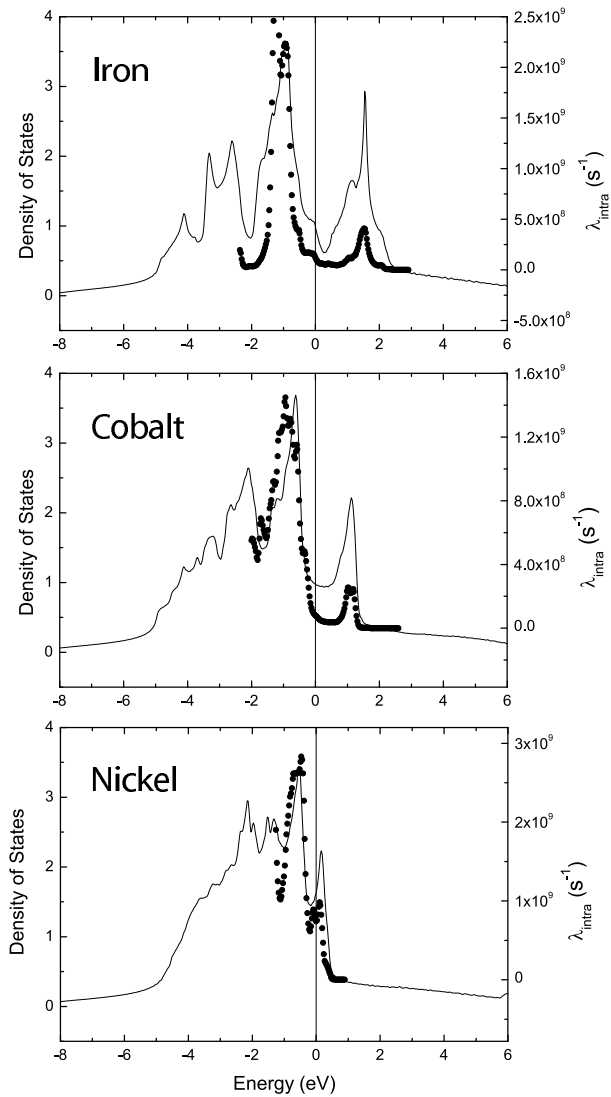}
\caption{Intraband damping rate versus Fermi level superimposed upon density of states.
A strong correlation between the intraband damping rate versus Fermi level ($\bullet$) and the 
density of states (solid curves) is observed.  Vertical black lines indicates true Fermi 
energy calculated by density functional theory.}
\label{dos.damping}
\end{figure}

For the intraband terms, the integral over the spectral functions is essentially proportional to
the density of states at the Fermi level.  Therefore, it appears reasonable to suspect that 
the intraband contribution to the damping rate of a given material should be roughly 
proportional to the density of states of that material at the Fermi level.  To test 
this claim numerically, we artificially varied the Fermi level of the metals within the
d-bands and calculated the intraband damping rate as a function of the Fermi level.  The
results of these calculations are superimposed on the calculated densities of states of the 
materials in Fig.~\ref{dos.damping}.  The correlation between the damping rates and the 
densities of states, while not exact, is certainly strong, indicating that increasing 
the density of states of a system at the Fermi level will generally increase the intraband contribution 
to damping.

The dependence of the interband terms on the spectral overlap is more complicated than that       
of the intraband terms.  The spectral overlap depends on the energy differences $\epsilon_m         
- \epsilon_n$, which can vary significantly between bands and over k-points.  When the              
scattering rate $\hbar/\tau$ is much less than these energy gaps the interband terms are            
proportional to the scattering rate.  However, this proportionality only holds at low               
scattering rates when the interband contribution is much less than the intraband              
contribution.  The proportionality breaks down at higher scattering rates when $\hbar/\tau$         
becomes comparable to the band gaps.  After this point the damping rate gradually plateaus          
with respect to the scattering rate.  Unfortunately, this complicated functional dependence         
of the spectral overlap on the scattering rate makes it difficult to obtain a simple                
description of the effect of the spectral overlap on the interband damping rate in terms of         
material parameters.

\subsection{Torque matrix elements}

The damping rate also depends on the square of the torque matrix elements.  A goal 
of doping is often to modify the effective spin-orbit coupling of a sample.  While doping does more than 
this, such as intoducing strong local scattering centers, it is nevertheless useful
to estimate the dependence of the matrix elements on the spin-orbit parameter 
$\xi$.  We begin with pure spin states $\psi^0_{n}$ and treat the spin-orbit interaction $V = \xi 
V^{\prime}$ as a perturbation.  The states can be expanded in powers of $\xi$ as
~
\begin{equation}
\psi_{n} = \psi_{n}^0 + \xi \psi_{n}^1 + \xi^2 \psi_n^2 + \dots \, .  
\end{equation}

\noindent The superscripts refer to the unperturbed wavefunction (0) and the additions ($i$) due to the 
perturbation to the $i$th order while the subscript $n$ is the band index, which includes the spin 
direction, up or down.  Since the torque operator also contains a factor of the spin-orbit parameter the 
matrix elements have terms in every order of $\xi$ beginning with the first order.  Therefore, the 
squared matrix elements have contributions of order $\xi^2$ and higher.

\begin{figure}
   \includegraphics[angle=0,width=8.0cm]{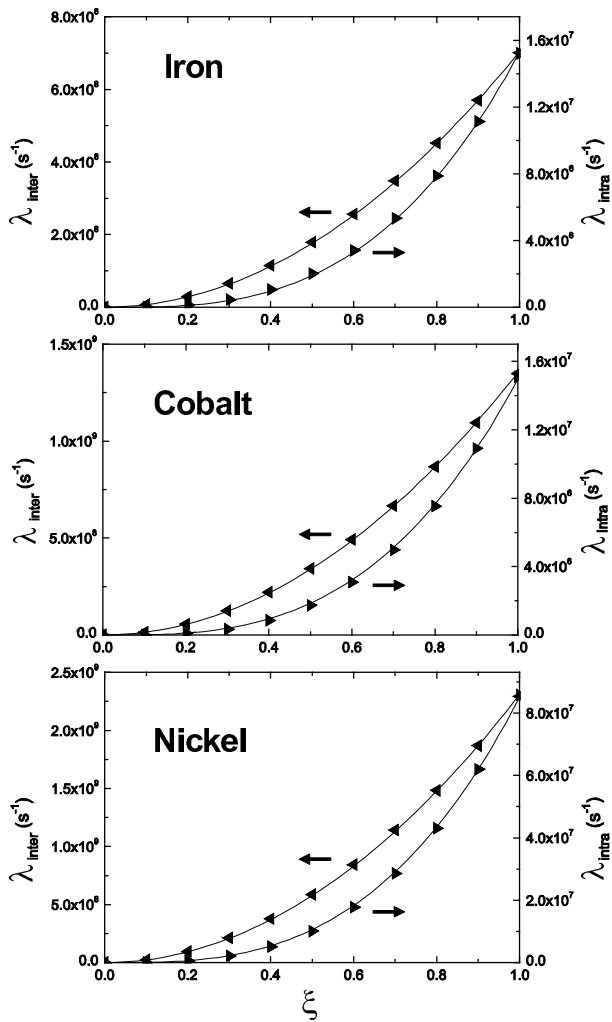}
\caption{$\xi$ dependence of intraband and interband damping rates.  Damping rates 
were calculated for a range of spin-orbit interaction strengths between off ($\xi=0$) and full strength 
($\xi=1$).  $\xi^2$ fits were made to the interband damping rates (left axes and 
$\blacktriangleleft$ symbols) and $\xi^3$ fits to the intraband rates (right axes and 
$\blacktriangleright$ symbols).}
\label{xi.fits}
\end{figure}

To determine the importance of these terms we artificially tune the spin-orbit interaction 
from zero to full strength, calculating the damping rate over this range.  We then fit 
the intraband and interband damping rates separately to polynomials.  In each material, this fitting 
showed that for the intraband terms the $\xi$ dependence of the damping rate was primarily 
third order, with smaller contributions from the second and fourth order terms.  Restricting 
the fit to only the third order term produced a very reasonable result, shown in Fig.~
(\ref{xi.fits}).  For the interband terms, polynomial fitting was dominated by the second 
order term, with all other powers contributing only negligibly.  The second order fit 
is shown in Fig.~(\ref{xi.fits}).

To understand the difference in the $\xi$ dependence of the intraband and interband 
contributions it is useful to define the torque operator
~
\begin{equation}
\Gamma^- = \xi( \ell^- \sigma^z - \ell^z \sigma^-) \, .
\label{torque}
\end{equation} 

\noindent The torque operator lowers the angular momentum of the state it acts on.  This
can be accomplished either by lowering the spin momentum $\ell^z \sigma^-$, a spin flip, or
lowering the orbital momentum $\ell^- \sigma^z$, an orbital excitation.  Therefore, both 
the intraband and interband contributions each have two sub-mechanism:
spin flips and orbital excitations. 

The second order terms for the intraband case are $\xi^2 |\bracket{\psi_n^0}{(\ell^z 
\sigma^- - \ell^- \sigma^z)}{\psi_n^0}|^2$.  Since the unperturbed states $\psi_n^0$ are 
pure spin states the spin flip part $\ell^z \sigma^-$ of the torque returns zero.  
Therefore, only the orbital excitations exist to lowest order in $\xi$, reducing the 
strength of the second order term in the intraband case.  However, the interband terms 
contain matrix elements between several states, some with the same spin direction, but 
others with opposite spin direction.  Therefore, both spin flips and orbital excitations 
contribute in second order to the interband contribution.

\section{Conclusions}

The breathing Fermi surface model has provided a simple and understandable 
effective field explanation of precession damping in metallic ferromagnets.  However, it is 
only applicable to very pure systems at low temperatures.  On the other hand, the torque 
correlation model accurately predicts damping rates of systems with imperfections from low 
temperatures to above room temperature.  The shortcoming of the torque correlation model is 
that it does not illuminate the phyiscal mechanisms responsible for damping.  We have pointed 
out that the breathing Fermi surface model accounts for only one of the two terms in the 
effective field.  By constructing an effective field with the previously studied breathing 
Fermi surface contribution and also the new bubbling effect we have shown that this simpler 
picture may be mapped onto the torque correlation model such that the breathing terms match 
the intraband contribution and the bubbling terms match the interband contribution.

Since there is considerable interest in understanding how to manipulate the damping rates of 
materials we investigated the dependence of the intraband and interband damping rates on 
both the spectral overlap integral and the torque matrix elements.  For the intraband terms, 
the spectral overlap is proportional to the density of states and we found a strong 
correlation between the intraband damping rate and the density of states of the material.  
The interband case is significantly complicated by the range of band gaps present in 
materials.  No simple relation was found between the strength or scattering rate dependence 
of the interband terms and common material parameters.  The importance of the torque 
matrix elements to the damping rates was characterized through their dependence on the 
spin-orbit parameter.  The intraband damping rates were found to vary as the spin-orbit 
parameter cubed while the interband damping rates went as the spin-orbit parameter squared.  
This difference was explained by noting that the torque operator changes the angular 
momentum of states either through spin flips, or by changing their orbital angular 
momentum.  Spin-flip excitations do not occur to second order in $\xi$ for the intraband 
terms, but do contribute at second order for the interband terms.

It is desirable to understand the relative differences in damping rates amoung various 
materials, such as why the damping rate for nickel is higher than that for cobalt and iron.
We have shown that the relative damping rates of these materials depend in part on the 
differences of their densities of states and spin-orbit coupling strengths.  However, they 
also depend in an intricate way on the energy gap spectra of each metal.  For the interband 
terms the dependence on the gap spectrum enters through the spectral overlap integral.  For 
the intraband terms the energy gaps appear in the denominators of the matrix elements.  
Therefore, states with very small splittings can dominate the k-space convolution. The 
abundance of such states in nickel appears to contribute to the larger damping rate in 
this material \cite{Kambersky:2007}.

Doping is a common technique for modifying damping rates.  Doping has a number of 
consequences on a sample and these effects vary with the method of doping.  Dopants can 
increase the electron-lattice scattering rate, introduce magnetic inhomogeneities that act 
as local scattering centers, alter the density of states, and change the effective 
spin-orbit parameter.  We have investigated the consequences of modifying the densities of 
states and spin-orbit parameter on the damping rate, and previously demonstrated the 
scattering rate dependence of the damping rate; however, it is not clear what new damping 
mechanisms arise when rare-earth elements are added to a transition metal host.

This work was supported in part by the Office of Naval Research through grant
N00014-03-1-0692 and through grant N00014-06-1-1016.


\bibliographystyle{apsrev}

\end{document}